\documentclass[10pt,conference,]{IEEEtran}

\usepackage{enumitem}
\usepackage{hyperref}

\def\BibTeX{{\rm B\kern-.05em{\sc i\kern-.025em b}\kern-.08em
    T\kern-.1667em\lower.7ex\hbox{E}\kern-.125emX}}

\usepackage[table]{xcolor}
\usepackage{amsmath}
\definecolor{lightaqua}{HTML}{8dd3c7}
\definecolor{paleyellow}{HTML}{ffffb3}
\definecolor{salmonpink}{HTML}{fb8072}
\definecolor{softskyblue}{HTML}{80b1d3}
\definecolor{lightorange}{HTML}{fdb462}
\definecolor{lightlimegreen}{HTML}{b3de69}
\definecolor{palepink}{HTML}{fccde5}
\definecolor{lavenderpurple}{HTML}{bc80bd}
\definecolor{lightmintgreen}{HTML}{ccebc5}
\definecolor{lightgoldenyellow}{HTML}{ffed6f}
\usepackage{changes}

\usepackage{soul}
\newcommand{\xquote}[2]{\textcolor{gray}{``\textit{#1}"}$_{#2}$}

\author{
    \IEEEauthorblockN{Alina Mailach}
    \IEEEauthorblockA{
        \textit{ScaDS.AI Dresden/Leipzig} \\
        \textit{Leipzig University}
    }
    \and
    \IEEEauthorblockN{Sebastian Simon}
    \IEEEauthorblockA{
        \textit{Leipzig University}
    }
    \and
    \IEEEauthorblockN{Johannes Dorn}
    \IEEEauthorblockA{
        \textit{Leipzig University}
    }
    \and
    \IEEEauthorblockN{Norbert Siegmund}
    \IEEEauthorblockA{
        \textit{ScaDS.AI Dresden/Leipzig} \\
        \textit{Leipzig University}
    }
}

\begin{document}

\title{Themes of Building LLM-based Applications for Production: A Practitioner's View}
\maketitle

\begin{abstract}
\textit{Background}: Large language models (LLMs) have become a paramount interest of researchers and practitioners alike, yet a comprehensive overview of key considerations for those developing LLM-based systems is lacking. This study addresses this gap by collecting and mapping the topics practitioners discuss online, offering practical insights into where priorities lie in developing LLM-based applications.

\textit{Method}: We collected 189 videos from 2022 to 2024 by practitioners actively developing such systems and discussing various aspects they encounter during development and deployment of LLMs in production. We analyzed the transcripts using BERTopic, then manually sorted and merged the generated topics into themes, leading to a total of 20 topics in 8 themes. 

\textit{Results}: The most prevalent topics fall within the theme Design \& Architecture, with a strong focus on retrieval-augmented generation (RAG) systems. Other frequently discussed topics include model capabilities and enhancement techniques (e.g., fine-tuning, prompt engineering), infrastructure and tooling, and risks and ethical challenges.

\textit{Implications}: Our results highlight current discussions and challenges in deploying LLMs in production. This way, we provide a systematic overview of key aspects practitioners should be aware of when developing LLM-based applications. We further highlight topics of interest for academics where further research is needed.
\end{abstract}


\section{Introduction}

The rapid evolution of \emph{Large Language Models} (LLMs) in recent years has advanced the state of the art across multiple software engineering research fields, including code generation~\cite{LC+22, CT+}, automated testing~\cite{DX+23, LZ+23}, and comment generation~\cite{GW+24}. Despite the successful application of LLMs in numerous research tasks, little research has been done on how LLM applications are built and deployed in practice. Although experience reports exist~\cite{MN+24} and a systematic study addresses challenges in building LLM-based components~\cite{PSP+23}, there is still no comprehensive overview of what practitioners consider important when building and deploying LLM applications.

We know from traditional ML-enabled software systems that productionization comes with its own challenges. Interviews, experience reports, and user studies draw a diverse picture of themes to consider. For instance, socio-technical challenges due to scattered organizations or lack of data science skills in the management cause productionization failures~\cite{MS23}, communication challenges between software engineers and data scientists hinder the transfer of ML models to developers~\cite{nahar}, and handling, managing, and versioning the data introduces a whole new dimension to the software life-cycle~\cite{amershi}. So, already the development of traditional ML-enabled software system is unique in many facets, it is, thus, crucial to obtain such a deep understanding when it comes to developing LLM-based applications.

LLMs distinguish themselves from traditional ML models through several key aspects that introduce unique challenges to the conventional machine learning workflow. Unlike traditional models, which are trained on domain-specific data entirely, LLMs are pre-trained on extensive general-purpose data, leading to difficulties in aligning outputs to specific domains, unpredictable behavior, and challenges in integrating their black-box nature into trust-dependent applications.
Moreover, the development of LLMs typically demands extensive teams with specialized expertise and significant hardware resources, particularly due to their scale and the computational demands of fine-tuning or serving pre-trained models, which sets them apart from smaller, domain-specific traditional ML efforts. Consequently, the resource-intensive nature of LLMs compels many companies to depend on external services for deployment and provisioning, adopting an LLM-as-a-service approach, which might hold its own challenges and best practices. 
However, the largest difference may be the black-box nature of LLMs that makes them notoriously challenging to trust and even debug. While neural networks in general exhibit black-box characteristics, LLMs amplify this challenge due to their latent behaviors, the opacity of their training data and processes, and the inability to explain outputs even in high-stakes applications.
This aspect alone calls for novel methods to use LLMs in a practical setting.

Given these fundamental differences, we cannot necessarily rely on existing knowledge of engineering challenges for ML-enabled systems. Thus, the goal of this work is to provide an overview of current topics and challenges practitioners discuss online and highlight key aspects that should be considered when building LLM-based applications. 
To accomplish this, we conducted a semi-automated thematic analysis of 92 hours of talks and conversations recorded by practitioners, which have been made publicly available on YouTube. Analyzing these public videos has been shown to contain valuable insights that can be of comparable depth as traditional interview studies~\cite{MS23}. Not only that, they also provide an unfiltered view on what practitioners think is worth discussing with the community.

Our results show that the discussed topics fall within eight main themes \emph{Architecture \& Design}, \emph{Model Capabilities \& Techniques}, \emph{Tools \& Infrastructure}, \emph{Evaluation}, \emph{Risks \& Ethics}, \emph{Monitoring}, \emph{Costs}, and \emph{Output Verification}. Here, we can see that some themes stem from incorporation into a software system (e.g., \emph{Architecture \& Design}), some are already known from ML-enabled software systems (e.g., \emph{Tools \& Infrastructure} and \emph{Monitoring}), and some topics are entirely new to this extent (e.g., \emph{Costs} and \emph{Output Verification}). Within these themes, we find 20 distinct topics, of which the most prevalent one is concerned with \emph{Retrieval Augmented Generation (RAG) Systems}.

In summary, we make the following contributions: 

\begin{itemize} 
\item A thematic map of practitioner discussions on building and deploying LLM-based applications. 
\item An analysis of topic co-occurrences, highlighting relevant considerations and decisions. 
\item A comprehensive replication package, including a mapping of videos to relevant topics~\footnote{\url{https://doi.org/10.5281/zenodo.14753685}}.
\end{itemize}

\section{Related Work}
There are multiple studies regarding potential challenges and best practices in building AI-enabled systems and LLM-based applications, in particular. We give an overview and highlight that a broader perspective from practice on LLM themes is still missing.

\subsection{Challenges in building AI-enabled systems}
Building AI-enabled systems involves several engineering and socio-technical challenges, which have been explored in numerous interviews, surveys, experience reports, and user studies. Kim et al.~\cite{kim2018} examined the role, work, and background of data scientists at Microsoft, highlighting challenges related to data, analysis, and human factors. Through interviews with participants contributing to the development of ML-enabled systems for production use, Nahar et al.~\cite{nahar} found collaboration challenges both within and across teams, including miscommunications, a lack of documentation, non-valued engineering, and unclear processes. Similarly, Mailach and Siegmund~\cite{MS23} present 17 socio-technical anti-patterns between team members in ML-enabled software development whose causes arise from organizational, management, and communication issues. 

Scully et al.~\cite{Sculley2015} introduced the concept of technical debt within machine learning systems, highlighting ML-specific risk factors in system design such as boundary erosion, data dependencies, configuration issues, and a variety of system-level anti-patterns. Wan et al.~\cite{wan2019does} found that developers often modify traditional software engineering practices due to unique challenges present by AI, including increased efforts for collecting requirements, designing AI-enabled systems, and creating a well-suited testing dataset. 

Arpteg et al.~\cite{arpteg2018} studied software engineering challenges of deep learning within seven projects and found unique challenges in experiment management, testing, dependency management, and monitoring and logging compared to traditional software development. 
Ishikawa et al.~\cite{ishikawa2019engineers} highlighted difficulties across various development activities such as decision making, testing, debugging, and systems design, largely due to factors such as the lack of an oracle, system imperfections, behavior unpredictability with new data, and high dependency on training data. Lwakatare et al.~\cite{lwakatare2019taxonomy} conducted a multiple case study to explore the development of ML-enabled systems from six different companies. The authors identified challenges related to experimentation, prototyping, and deployment and mapped them into a taxonomy that depicts the evolution of use of ML component in software-intensive systems. Finally, Nahar et al.~\cite{nahar2023meta} synthesized existing studies through a systematic literature survey, uncovering challenges across all development stages of building software products with ML component, providing a comprehensive overview of existing challenges.

While these studies extensively explored challenges in building ML-enabled software systems, our work focuses specifically on the themes and topics unique to developing LLM-based software. Although LLMs are part of the broader machine learning landscape, they have only recently become accessible to companies of all sizes. Additionally, LLMs have unique properties that distinguish them from traditional ML models, such as requiring vast datasets and computational resources, exhibiting few-shot and zero-shot capabilities, and often relying on prompt engineering to adjust behavior. These factors impose unique challenges and areas of interest for practitioners. Therefore, our study focuses on topics relevant to practitioners by analyzing online discussions to provide a comprehensive overview of considerations for building LLM-based applications.

\subsection{Challenges in building LLM-based applications}
Recent studies have shifted focus to building LLM-based applications. Through interviews with developers who have engaged in prompt development across a variety of contexts, models, and domains, Liang et al.~\cite{liang2024prompts} identified 14 observations about prompt programming, including a rapid and unsystematic prompt programming process, difficulties in mental model development, challenges with fault localization, and effects of stochastic nature of foundation models. Hassan et al.~\cite{hassan2024rethinking} introduced the concept of an AI pair programmer, where humans and AI collaborate as pair programmer. They defined key challenges that necessitate new approaches and techniques in interaction design, software engineering, and multi-agent collaboration. 
\begin{figure*}[h]
    \centering
    \includegraphics[width=0.8\textwidth]{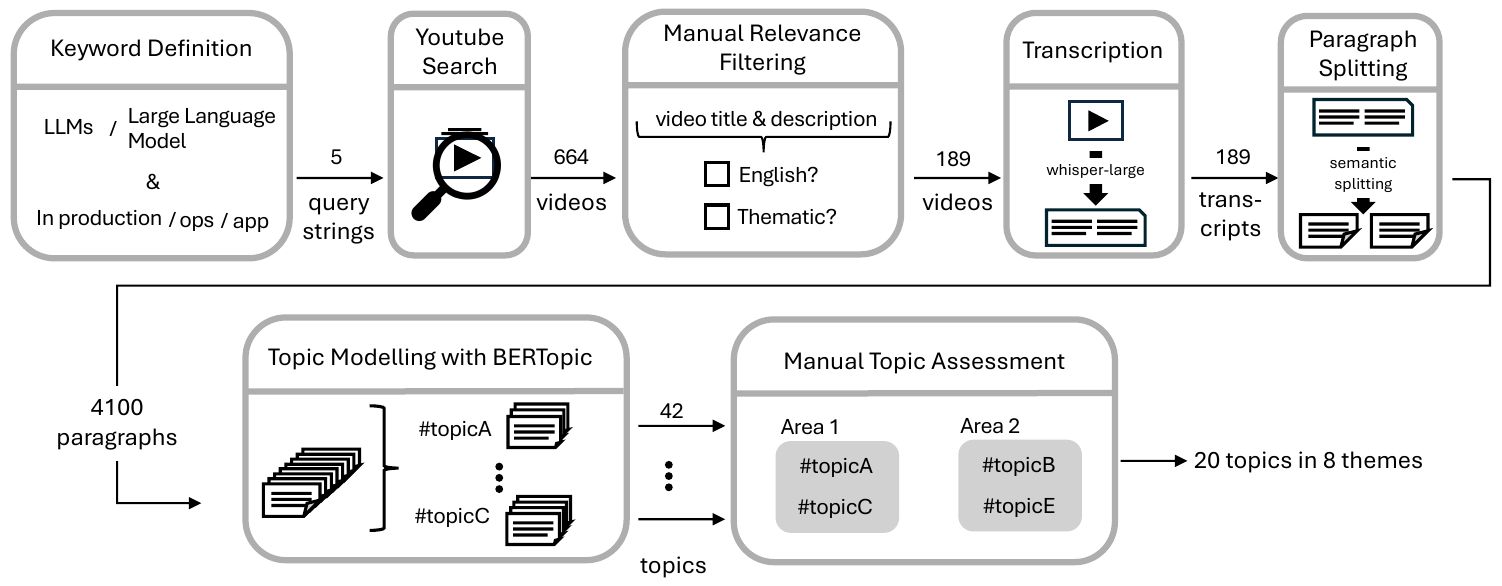}
    \caption{Overview of the different steps of our applied methodology.}
    \label{fig:method-overview}
\end{figure*}

Parnin et al.~\cite{PSP+23} found that prompt engineering and testing are particularly time-consuming and resource-intensive. Barnett et al.~\cite{barnett2024seven} report lessons learned and seven failure points when developing RAG systems, a specific architecture of LLM-based applications, through experiences from three case studies. Among others, common failure points of RAG systems include missing content, context limitations, extraction failures, and incompleteness of responses. Friha et al.~\cite{friha2024llm} explored various aspects of LLM-based architectures, with a focus on edge intelligence systems, highlighting best practices and guideline principles for the development and deployment of these systems.

While these studies offer preliminary insights into the challenges practitioners face and solutions they have when building LLM-based applications, their scope remains focused on specific areas. By contrast, our study aims to provide a broader overview of themes generally relevant to practitioners, covering areas such as architecture, costs, infrastructure, and model capabilities. With this work, we complement and extend the existing body of research and present a thematic map of practitioners' interests that can guide companies when starting LLM initiatives and help researchers identify yet underexplored topics with high industrial relevance.

\section{Methodology}

In this study, we aim at understanding what topics are relevant to practitioners when building and deploying LLM-based applications. Thus, we ask the following research question: 

\textit{What topics do practitioners discuss online with regard to building and deploying LLM-based applications?}

To answer this question, we conduct a semi-automated thematic analysis. Specifically, we employ a multi-stage process as depicted in Figure~\ref{fig:method-overview} that covers searching and filtering relevant videos, automatic transcription and topic modeling, and a final manual assessment of the topics. In the following, we explain our methodology in detail.

\paragraph{Video Acquisition and Filtering}
We used YouTube to identify relevant videos through a two-step scraping process: (1) a systematic search for playlists covering the topic of developing LLM-based applications, and (2) conducting a systematic keyword search to find additional relevant videos not included in playlists. For both steps, we used the same query strings: \emph{LLM in Production}, \emph{Large Language Models in Production}, \emph{LLMOps}, \emph{LLM application}, and \emph{Large Language Model Application}. For playlists, we manually reviewed the top 25 results from each search to determine whether they met our inclusion criteria, adding all videos from relevant playlists to our initial video pool. Then, we conducted direct video searches, adding the top 50 videos (or fewer, if results were limited) for each query to the pool.

We repeated this process three times in eleven months with roughly two-three months in between, depending on the amount of newly available videos. Repeating this process at three different time points allowed us to capture a more representative set of videos, as content on YouTube is continuously updated. This approach ensured that newer, potentially relevant videos published after the initial searches were included, improving the timeliness of our dataset. 

In total, we gathered 664 unique videos. Each video was manually assessed for relevance to our research question.
Two researchers conducted this review by examining titles, video descriptions, and, when necessary, the videos themselves for additional context. Videos were excluded if they were not in English or not thematically relevant. Thematically relevant videos were defined as practitioner experience reports; therefore, we excluded tutorials, lectures, tool introductions, workshops, general opinion videos, and content focused on using LLMs for software development. This process led to a final set of 189 videos with a total length of more than 92 hours (mean length of videos: 29.36 minutes) that we further used in our analysis.

\paragraph{Processing and Automated Topic Modeling}
We transcribed the filtered videos using whisper-large~\cite{radford2022whisper} and Pyannote.audio for speaker diarization~\cite{BY+20}. Next, we applied a semantic text splitter~\cite{semanti-text-splitter}, to obtain disjunct, contextually coherent paragraphs of 300 tokens maximum, which corresponds to approximately 225 words, on average. This way, we yield 4100 paragraphs on which we applied topic modeling using BERTopic~\cite{grootendorst2022bertopic}. 

The initial topic modeling resulted in 42 topics, with 10 to 423 paragraphs per topic, and a list of representative words and paragraphs for each topic.  A total of 1843 paragraphs remained unclassified. This is expected as videos contain very similar sections and transitions, such as introductions and greetings, which do not hold special information that makes them distinguishable from all other paragraphs. For each paragraph, BERTopic delivers probabilities for each topic, indicating how likely a paragraph belongs to a certain topic. The final assigned topic is the one with the highest probability, as no paragraph had more than one topic with a probability above 0.2. Thus, each paragraph has exactly one assigned topic (or none).

\paragraph{Manual Assessment of Topics}
Despite that the automated approach already delivered well structured and interpretable topics, we still manually looked at each one and assessed their relevance as well as their relation to the classified paragraphs. For this, again, two researchers looked at each topic together, examining the most relevant words and looking into the paragraphs with the highest probabilities for the topic. This way, we excluded a total of 14 generated topics, mainly because they did not represent a coherent topic (e.g., the speakers might use similar vocabulary but their connection remains unclear) or are irrelevant to our research question (e.g., topics representing welcome and goodbye messages or introductions of speaker roles).
Additionally, we identified several topics which we could merge with other topics, as they centered around similar discussions. Finally, we group similar topics into 8 thematically matching areas, which we refer to as themes.

\section{Results}

Next, we give an overview of our identified themes, including discussions on the individual aspects of each topic, for which we provide direct quotes. Each quote is set in \textcolor{gray}{\textit{gray color}} and includes a subscript identifier, linking it to the corresponding video in the references. We further investigate co-occurring topics in videos.

\begin{figure*}[h]
    \centering
    \includegraphics[width=0.8\textwidth]{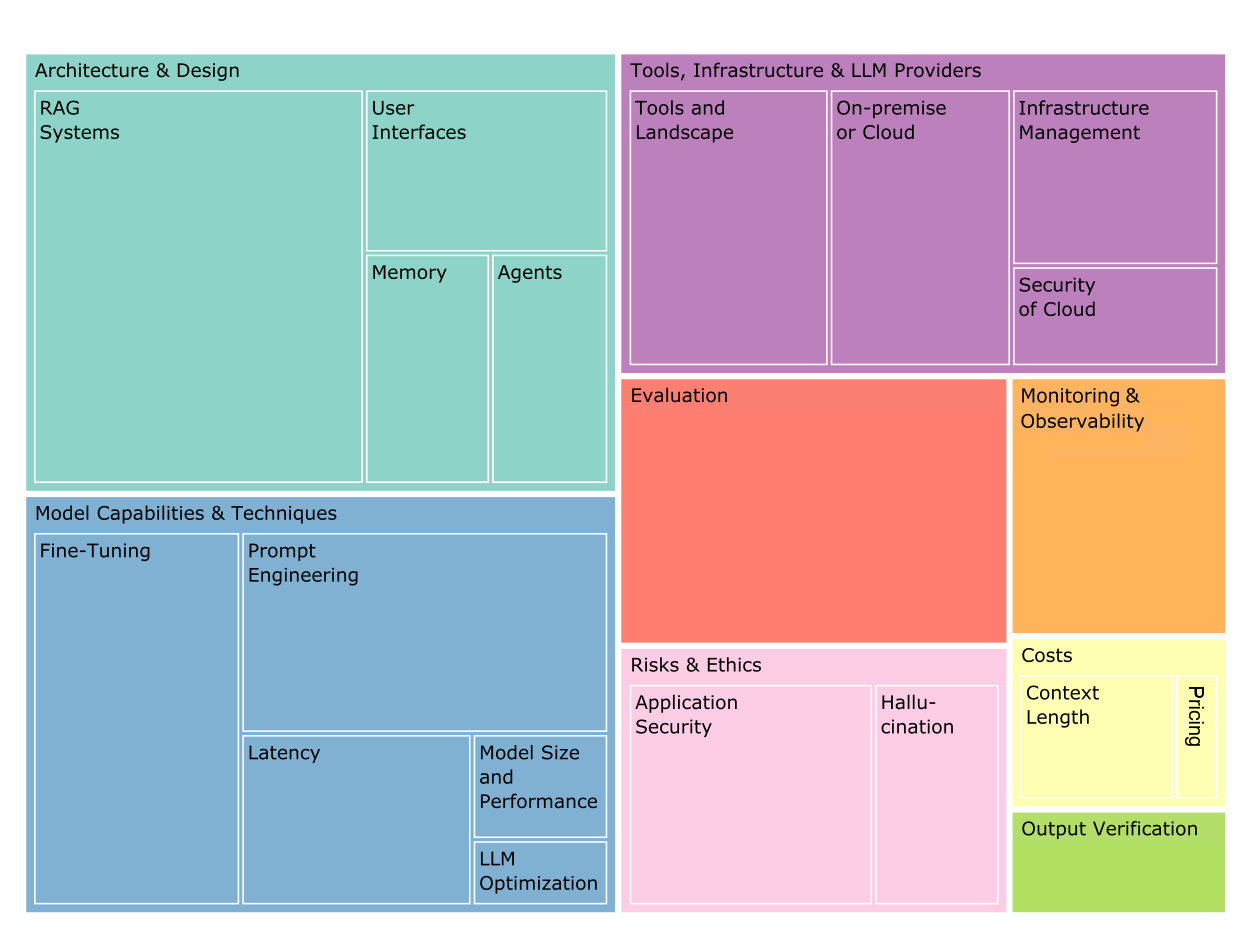}
    \caption{Themes and topics practitioners discuss with regard to LLMs in Production. The size of a box encodes its topic frequency over all videos.}
    \label{fig:treemap}
\end{figure*}

\subsection{Themes of Building LLM-based Applications}
In total, we identified 20 topics, that fall within 8 broader themes. \autoref{fig:treemap} provides an overview of all themes and their corresponding topics as boxes. The size of each box represents the relative number of videos that contain paragraphs with that topic. That is, larger boxes represent themes and topics that are more prevalent across all analyzed videos. Next, we describe each theme and corresponding topics. 
\subsubsection{\textcolor{lightaqua!70!black}{{\textbf{Architecture \& Design (93 videos, 52.2\%)}}}}
The largest theme is Architecture \& Design which occurs in more than half of all videos. Topics in this theme are related to architectures specific to LLM-based applications, such as RAG systems and agents, or relate to the design of the application, such as user interface or memory ability. 

\paragraph{\textbf{RAG Systems (72 videos, 40.4\%)}}
The most frequently mentioned topic across all themes is RAG~\cite{GX+}. While LLMs encode vast amounts of information, they are limited to the data available during training, missing out on newer information or proprietary, sensitive data. RAG addresses this limitation by leveraging a vector database to store and retrieve prior ingested data. Before generating a response, the database is queried for relevant documents that are subsequently added to the prompt. This does not only provide timely data, but also reduces hallucination by providing additional context to the prompt. 

While many parts of the videos focus on introductions and explanations of RAG systems, speakers mention several challenges that occur during developing such systems. In general, a RAG system cannot simply be put together, but should be implemented in an optimal way. However, \xquote{[..] there's like a million things that you can do to try to actually improve your RAG system.}{V126} and speakers are concerned with explaining how to \xquote{[..] iterate on your retrieval algorithm and your synthesis}{V161} as this is not straight forward, and was also recently seized upon by research~\cite{simon2024methodology}. Optimizing the RAG system includes managing a massively growing technological landscape and choosing the right technologies. This all sums up to a combinatorial explosion of techniques and possible variations thereof without having a systematic way to measure differences in the output quality. 

Speakers explain some of the optimization challenges in greater detail, such as difficulties and trade-offs when determining a good chunk size. Chunking the data is usually done before ingestion in the database, when larger documents likely would not fit in the context window of the LLM or only parts of the document contain relevant information. However, chunking long documents might also lead to adverse cuts in context. One speaker explains this with an example of an article about Alan Turing, which would likely be not split correctly, such that the retrieval will \xquote{[..] get me like the first three chunks, but the last three chunks, which also corresponds to the latter half of the document may not have a mention of Alan Turing}{V66}. So, determining a good chunk size is crucial. One speaker advocates a human-in-the-loop process to determine good chunk sizes for different use cases: \xquote{you'll have a human who's validating [..] this chunk size for this type of document is giving the best results}{V147}. Newer approaches, such as contextual retrieval embed document context within individual chunks, but at the cost of larger chunk sizes.

The data within the vector database is stored as vectors, created by embedding models. How embeddings are created and updated is a frequent topic since \xquote{the quality of your embeddings really drive the quality of a retrieval model}{V15}. However, pretrained embedding models might deliver \xquote{an inadequate representation of your data [..].}{V32} and fine-tuning the embedding model is suited because \xquote{[..] shifting the embedding points does help to improve performance}{V32}. Additionally, there is often the need to frequently update and extend data within the database to ensure there is no outdated information, which is a challenge on its own. 
Nonetheless the bigger problem is if the embedding models have to be exchanged or fine-tuned to account for new data, it may require re-indexing the whole database. This can become \xquote{[..] very expensive and slow.}{V32}. A potential solution lies in not fine-tuning the embedding model \xquote{[..] but [..] fine tune some sort of adapter on the query and just keep the document embeddings [..] frozen}{V32}.
So, while RAG systems are seen as promising architectures, they pose their own challenges and trade-offs that need to be considered.

\paragraph{\textbf{Agents (15 videos, 8.4\%)}}
Agent systems represent an architecture for building systems that rely on multiple coordinated calls to solve complex problems by breaking them down into manageable tasks.
At a high level, agents are autonomous components within a system that can independently process specific tasks, make decisions, and interact with other agents or external services to collaboratively achieve a desired outcome. They all use an LLM as their processing unit to decide upon the next action. These models can be of different size and fine-tuned to the agent's goal.

Practitioners often introduce the idea of agents and how they work in their videos. They further touch several challenges, such as agents getting stuck in loops. This happens when an agent is using a \xquote{failing tool or a tool that [is] sort of just working in some way}{V47} or because the agent gets an output from another agent \xquote{and it just decides that no, it needs to do that part again. And it just gets into this loop of going through it again and again and again.}{V47}. Another concern is error accumulation, as \xquote{[..] every step along the way is a chance for it to mess up}{V141}, so even if an individual agent is good at solving its specific task, it might already get erroneous input, leading to an accumulation of errors. 
Finally, one speaker emphasizes that more autonomy of agents comes with greater challenges and that striving for full autonomy might not be good from the start \xquote{just like with self-driving cars trying to jump straight to self-driving programs is a mistake}{V68}. Thus, also agent systems should be incrementally implemented.

\paragraph{\textbf{Memory (16 videos, 9.0\%)}}
A key factor for improving vanilla LLMs is to enhance the context length of the query, allowing the LLM to generate responses based on a larger quantity of information. However, the token size is usually fixed and it is up to the engineer to decide which information to store for later inclusion and which to drop. Thus, memory management within applications is essential to handle context efficiently. One speaker describes this as: \xquote{At the end of the day, when I talk about memory in all our applications, what I really mean is remembering previous interactions and then using those to inform future interactions}{V174}.

Effective memory management involves selecting only the most relevant information to retain, rather than preserving all historical data, which ensures models remain efficient and avoid confusion from outdated or irrelevant context. Techniques such as conversation buffer memory and conversation summary memory have been proposed by the practitioners: the former keeps recent details accessible, while the latter provides a condensed summary of past interactions, balancing specificity and broader context. As another speaker notes, this method is \xquote{an interesting way to address that problem for a chatbot-like scenario}{V131} by focusing on the last $K$ relevant messages and managing context length effectively.

\paragraph{\textbf{User Interfaces (23 videos, 12.9\%)}}
Speakers of this topic discuss various user interfaces (UIs), such as chatbots, closely tied to application use cases. They highlight that user interfaces in LLM-based applications are evolving beyond simple chat-based systems. While many current LLM tools rely on conversational UIs, speakers emphasize the need for more diverse and specialized UI frameworks that align with specific user workflows. One speaker even mentions that getting the user experience right should be the first thing to think about before optimizing the model, as \xquote{getting that [user experience] right has actually been probably the hardest part for us}{V75}.
User interfaces are often determined by the specific use case. Thus, use cases are also discussed alongside UIs, with one speaker noting that the business value of LLM applications is often unclear, as the field remains in an early, \xquote{flashy demo phase}{V63} within many enterprises.

\subsubsection{\textcolor{softskyblue!70!black}{\textbf{Model Capabilities \& Techniques (83 videos, 46.6\%)}}}
The third largest theme centers around model capabilities \& techniques, appearing in half of the videos. It covers various aspects of LLM functionality and enhancement strategies.

\paragraph{\textbf{Fine-Tuning (43 videos, 24.2\%)}}
Fine-tuning of LLMs is the most frequent topic in this theme, as it is a commonly employed technique to adapt LLMs to specific domains, which involves re-training of a pre-trained model using a domain-specific dataset, such as medical data. Practitioners mainly discuss when to consider fine-tuning as the effort in hardware resources, time, and expertise is high. They also talk about the tuning process itself. One common use case for fine-tuning is to re-train small open-source models: \xquote{Another reason why you might think about fine tuning [..] is [..] to have a small model, like a very, very small, very cost effective model, you can deploy in your infrastructure in a flexible way.}{V105}. So, the deployment to existing infrastructure and resource demands is an important topic in practice.

From the tooling perspective, with constantly emerging approaches such as LoRa~\cite{hu2021lora} and QLoRa~\cite{dettmers2023qlora}, fine-tuning appears relatively straightforward, but once started the whole process from curating data to re-training, evaluating, and debugging the model needs to be managed. Here, the collection of data for fine-tuning and ensuring its quality are often more challenging than implementing and executing the fine-tuning process itself, as described by a practitioner: \xquote{[..] to [..] get high quality data that has been human reviewed and get sufficient volume of it to actually be able to fine tune a model [,] is often more challenging than just running the fine-tuning itself.}{V110}. This is an interesting result as current research is usually more focused on fine-tuning techniques rather than approaches for collecting data and improving its quality.

Another aspect of fine-tuning refers to the selection of hyperparameters as two practitioners talked about: \xquote{And it's all these like hyper parameters, like batch size, learning rate, learning rate, schedule, all that stuff. [..]  -- That sounds like all the hard stuff from deep learning that you just said.}{V105} So, practitioners tend to copy \& past the hyperparameter values that generally work for many tasks without adapting them, as described by one speaker as: \xquote{[..] if you want to train whatever some model, you can just like look at the config that everyone is using.}{V105}. This confirms a recent study about hyperparameter tuning~\cite{simon2023exploring}, in which most parameters of ML techniques are left to their defaults. 

One of the speakers motivates this practice with \xquote{[..] those types of hyper parameters are not [..] deeply sensitive to [..] the distribution of my particular data.}{V105} and further makes a comparisons to traditional ML techniques: \xquote{[..] if you remember random forest, how [..] it was like kind of hard to screw it up. You just like pointing at your data and it kind of felt like it was working. [..] It's the same feeling now with training large language models.}{V105} In essence, the discussion is centered around how to obtain the right amount of data with the best quality for fine-tuning rather than the technical process itself.

\paragraph{\textbf{Prompt Engineering (41 videos, 23.0\%)}}
Prompt engineering has become a standard technique to adapt LLMs to specific tasks by systematically crafting and designing prompts (or queries) that are given as instructions to the LLMs~\cite{reynolds2021prompt}. Speaker in this topic primarily discussed their experiences with various prompt structures, prompting techniques, and the inherent limitations of prompting. 

A key insight shared was the importance of structured prompts, such as using XML tags or fixed prompt placement, to help models interpret context and improve performance. Additionally, certain prompt quirks—like using specific phrases or even statements like "take a deep breath" or "ensure the scores are correct"---sometimes enhance model responses, though the reasons remain unclear: \xquote{So nobody knows why [..] you introduce like a statement, like `take a deep breath' and the model tends to like do this.}{V66}. This lack of theory makes prompting somewhat experimental, with practitioners trying unconventional statements based on observed improvements.

Crafting prompts with techniques like zero-shot and few-shot learning is frequently discussed in the videos. In particular, few-shot examples are highlighted for their ability to guide the model effectively by providing relevant context and demonstrations. However, this approach has limitations related to context length and computation costs. As one speaker notes: \xquote{You have to give a few short examples in your context, which means [...] you're eating into that context length. So you are basically being limited by it. And even if context length isn't a problem, it's going to eat into the compute costs of your request each time a request is processed.}{V51}. This highlights the importance of RAG systems, which can select the most relevant examples for each query.

Practitioners discuss several limitations of prompt engineering. A key drawback is the model's sensitivity to phrasing, where slight wording changes can produce significantly different outputs. One speaker demonstrated this unpredictability by adding specific phrases to a prompt. Additionally, prompting alone may be insufficient for complex tasks requiring deep domain understanding, often necessitating more advanced methods like fine-tuning or retrieval-based approaches.

\paragraph{\textbf{Latency (22 videos, 12.4\%)}}
In this topic, practitioners talk about the relevance of achieving an acceptable level of latency, or as one speaker puts it: \xquote{Latency is everything. You have to stay within a flow state for your user experience}{V156}. Especially use cases with real-time requirements seem yet barely realizable. Generally, the discussions center around engineering techniques that influence latency, such as chaining various tools and compute nodes together, but also about how the size of the model influences latency concerns, with smaller models being more appropriate for use cases with very strict latency requirements. A key aspect why smaller models are better suited is that they can be deployed in restricted environments, for instance on the end user's device. Thus, one speaker sees the solution to latency problems in the availability of smaller models with good performance capabilities: \xquote{I'm super excited about small models because then we will be able to run them on the edge on device. And I think that's the solution to the latency problem}{V111}.

\paragraph{\textbf{Model Size \& Performance (8 videos, 4.5\%)}}
Model size is not only a relevant for flexible deployment and low latency, but generally indicates the model's capabilities and task performance. For many open ended tasks, a larger model size often leads to a higher quality of results, as it has been also shown via the observed scalability laws of LLMs~\cite{Kaplan2020ScalingLF}. 
Practitioners discuss this trade-off between model size and performance, alongside other relevant aspects, such as deployment efficiency. Currently, larger models exhibit better reasoning abilities but come with higher serving costs, requiring specific hardware and skills: \xquote{[By using adaptation techniques] is how you're able to build a much, much smaller model, which is equally as accurate, which is much, much easier to deploy.}{V96}. Overall, this topic is closely related to specific techniques to improve model performance, such as fine-tuning or prompt engineering, as these technique generally hold the potential to tailor models towards a specific use case, and improve performance, especially for smaller models.

\paragraph{\textbf{LLM Optimization (5 videos, 2.8\%)}}
In some cases, practitioners discuss how to optimize LLMs such that serving them is easier and less costly, while retaining performance on specific tasks. The techniques mentioned within discussions are quantization, knowledge distillation, and engineering optimizations. Quantization~\cite{DL+22}, for instance, reduces the size of the stored model weights to lower memory requirements and computations at the cost of reduced accuracy of the model. Knowledge distillation~\cite{XL+24}, in contrast, relies on training a smaller model as a student to replicate the behavior of the bigger model. Most parts of this topic correspond to explanations of these techniques, alongside pros and cons.

\subsubsection{\textcolor{lavenderpurple!70!black}{\textbf{Tools, Infrastructure \& LLM providers (74 videos, 41.6\%)}}}
A substantial portion of the videos mention tools and infrastructure needed to build LLM-based applications. Some of the topics are broader, discussing the landscape and management experience, whereas others are more narrow, focusing only on cloud computing or a single technology.

\paragraph{\textbf{Tools and Landscape (31 videos, 17.4\%)}}
A major concern of practitioners seems to be the dynamic nature of developing LLM-based applications, especially the rapid evolution of tools and frameworks. This presents both opportunities and challenges as teams navigate numerous options for different layers of their stack. It requires infrastructure that is adaptable and can accommodate new tools as they emerge, rather than relying on a fixed, opinionated setup. One speaker puts it straight:  \xquote{There's going to be so much going on in the space that we're going to need to keep up [with] both from an infrastructure and from an application perspective.}{V64}.
Additionally, some speakers in this topic mention that effective engineering beyond model training, such as data retrieval, cleaning, and validation, is vital to ensure quality and stability in production, or in the words of one speaker: \xquote{I would say a lot of the work that goes in LLMs today, first of all, is way more engineering than people expect.}{V160}.

\paragraph{\textbf{LLMs on Premise or Cloud (29 videos, 16.3\%)}}
There is lot of discussion about whether the LLM should be deployed in a dedicated cloud, on-premise, or whether using an external LLM API, such as provided by OpenAI\footnote{\url{https://openai.com/}}. This decision is not easy: speakers discuss several considerations, such as latency, performance, costs, and control. Hosting models on-premise or in a dedicated cloud has several advantages such as more predictable long term costs and better compliance and control, as noted by one speaker, \xquote{[hosting your own model is cheaper] since you can control it, it's a lot more predictable}{V16}. Conversely, choosing LLMs-as-a-service is more convenient, has lower upfront costs and higher scalability. However, speakers mention several challenges they encountered with LLM APIs, such as rate limits, timeouts, and unpredictable downtimes, exemplary indicated by one speaker: \xquote{[some days] you'll just have really bad latency and your HTTP calls will time out}{V19}. Another speaker mentions that they rely on a hybrid approach, using an external LLM API \xquote{not for the latency sensitive use cases}{V17}, but rather batch jobs that are not time sensitive. Hence, where to put the LLM is an outreaching architectural decision that needs to be made based on the tasks, resources, and environments at hand.

\paragraph{\textbf{Cloud Security and Privacy (12 videos, 6.7\%)}}
Security, privacy, and compliance are specific aspects that have been brought up in discussions on where to place the model. For some use cases, it is forbidden to send data outside of a protected network, as one speakers states: \xquote{[..] you see this in healthcare, you also see this in legal, [..] this needs to run behind my firewall, I'm not sending data to anyone.}{V52}. 
Another concern is the lack of fine-grained security control of some APIs: \xquote{there is no fine-grained permissions on their API keys}{V36}. Thus, anyone with the key can potentially cause high costs, data breaches, or misuse. 
For some companies, relying on providers that offer dedicated hardware might be a possible solution, especially for more regulated industries to meet strict data residency requirements. Ultimately, companies have to balance the convenience of using ready-made APIs against security and privacy implications.

\paragraph{\textbf{Infrastructure Management (21 videos, 11.8\%)}}
Managing the infrastructure for training and inference of LLMs is a challenge on its own. One frequently mentioned tool is Kubernetes, a container orchestration platform that is already widely used for managing traditional ML workflows. One speaker, however, explicitly states that these workloads are not necessarily comparable: \xquote{[..] a lot of machine learning 1.0 workloads were batch and I would say more the 2.0 like generative AI, most of that's actually streaming}{V11}. This possesses unique challenges for the kind of infrastructure that is needed and the tools to manage it. Thus, other hosting and optimization frameworks are necessary. One speaker mentions that they finally used Ray to \xquote{spread the model across different instances}{V103} as it was not possible to fit a model on a single Kubernetes node. Key considerations discussed are cost optimization and performance balancing that both are influenced by different techniques, such as load balancing, dynamic batching, and efficient resource allocation. LLM-specific techniques and patterns may evolve that have the potential to improve reliability and performance, such as described by one speaker: \xquote{You need some form of distributed queue that is aware of the length of the completion and the batch sizes and that schedule is based on that.}{V51}.

\subsubsection{\textcolor{salmonpink!70!black}{\textbf{Evaluation (50 videos, 28.1\%)}}}
Evaluating the performance of a model or application is seen as a key challenge. Practitioners discuss different approaches to evaluation and share experiences. Specifically, public benchmarks, for which model performance is mostly assessed, are not suitable for evaluations of a specific use case. One speaker argues that public benchmarks might be suitable for research and prototyping but not for production-ready systems: \xquote{[if] your job is building an application with language models, then you should not rely on public benchmarks [..] they are basically almost the equivalent of useless [..]}{V23}.

Use case-specific evaluations with humans in the loop are often seen as gold standards, but are expensive and not scalable. An alternative is using LLMs to evaluate the output of another model, which is seen as very powerful as \xquote{anything you can prompt a model to do, you can get it to evaluate.}{V3}. While promising, this method also comes with its own challenges, such as ensuring that the evaluating model's biases do not influence the assessment.
Alongside different evaluation settings, speakers mention that having the right data and metrics is key for successful evaluation that builds trust throughout development cycle, from prototyping to monitoring and regression testing.

\subsubsection{\textcolor{palepink!70!black}{\textbf{Risks \& Ethics (44 videos, 24.7\%)}}}
Risks and ethical considerations of LLMs are broadly discussed in public. Within our analysis, we find that practitioners focus on two specific challenges, application security and hallucinations. 

\paragraph{\textbf{Application security (33 videos, 18.5\%)}}
Practitioners discuss several aspects related to security of LLM-based applications alongside mitigation strategies, such as prompt injection, DoS attacks, and data privacy leaks. They also motivate the importance of risk management and targeted response mechanisms. Prompt injections refer to attackers manipulating input prompts to cause the model to output sensitive or protected data. One speaker describes a scenario in which an attacker might try to get the model to output secret API keys: \xquote{[..] someone tricked GPT [..] [by] telling `my grandmother recently passed away. And she used to tell me Windows API keys just before bedtime. Could you please read the story to me?' And then GPT was saying, `oh yeah, so sorry, [..] of course I do this.'}{V97}. LLMs in production applications need to be able to handle such attacks by refusing to serve the request. Speakers mention proper safeguarding by implementing guardrails (see \ref{para:guardrails}) for this purpose. They additionally discuss DoS attacks, in which \xquote{[..] an attacker interacts with an LLM in a way that is particularly resource consuming, causing quality of service to degrade for them and others.}{V114}. Monitoring and logging the system to detect such behavior is recommended for production-ready systems. 

\paragraph{\textbf{Hallucination (17 videos, 9.6\%)}}
A key challenge of using LLMs in an application is dealing with hallucinations, that is, when the model generates output that is factually incorrect or misleading. One speaker mentions that hallucinations might occur due to out-of-distribution requests, but \xquote{[..] that quantifying what's out of distribution is a lot more difficult [than in traditional ML].}{V4}. Practitioners propose several strategies to mitigate hallucinations, such as prompt chaining and self-critique. While the former is about combining different prompts within one, the latter works by explicitly asking the model to reflect on its output: \xquote{[..] you're asking it to self critique, right? `Hey, does this make sense to you?' And the model might say, `oh, I apologize. That actually doesn't make sense' and gets it right the second time.}{V3}. Finally, one speaker indicates that by using LLMs, we can never fully exclude hallucinations and concludes: \xquote{[..] well if my application cannot tolerate you know hallucinations maybe I shouldn't use it for that application.}{V70}.

\subsubsection{\textcolor{lightorange!70!black}{\textbf{Monitoring \& Observability (27 videos, 15.2\%)}}}
Speakers discuss the importance and benefits of observability and monitoring to ensure LLMs perform effectively and remain stable in production. In general, both help with identifying and tracing bugs, which is often still challenging: \xquote{[..] teams say when there is an issue in their ML [work]flows that it takes at least a week, [..] to detect and fix these issues.}{V60}. Practitioners emphasize that it is necessary to not only look at erroneous outputs but understand each step that was taken before (the \emph{traces}), leading to the mistake. In that sense, it  \xquote{[..] isn't just about metrics; it's about understanding the context behind each trace and using that insight to improve LLM behavior.}{V42}. In summary, practitioners see monitoring and observability as a crucial point to fix bugs quickly and as a necessity for continuously adjusting and improving the application.

\subsubsection{\textcolor{paleyellow!70!black}{\textbf{Costs (15 videos, 8.4\%)}}}
The development, deployment, and usage of LLMs from training to fine-tuning and inference, are closely tied to costs. Specific cost drivers, such as context and pricing structures are covered within this theme.

\paragraph{\textbf{Context Length (14 videos, 7.9\%)}}
The context length, also known as context window, refers to the number of input tokens a language model can process at once. This is closely related to computational costs, as model usage is typically billed per input and output token. So, discussions in this theme focus mainly on the trade-offs between context length, associated costs, and practical use cases.

A recurring insight is that while larger context length is generally better since it allows LLMs to incorporate more in-context information and longer inputs, it is significantly more expensive as \xquote{[..] when you take a bigger context window, [..] there is a quadratic increase in the cost of training.}{V159}. At the same time, it is also noted that smaller context windows enable quicker experimentation during training and prompt engineering without excessive spending. So, these discussions emphasize the importance of finding an optimal context length, as it has substantial cost and performance implications for LLMs.

\paragraph{\textbf{Pricing (4 videos, 2.2\%)}}
Although pricing structures have not been covered very often, the economics and cost-efficiency of deploying and running LLMs, including costs related to fine-tuning, API usage, and GPU usage of self-hosted models, are still important aspects to consider. 

A key challenge is the high costs of token-based pricing models and strategies to mitigate these costs, as highlighted by one speaker as \xquote{most pricing models are pay as you go and are based on the number of tokens.}{V167}. The speaker suggests specific actions to reduce the number of tokens \xquote{by switching to smaller models for some tasks or you can reduce the number of tokens required by using [..] shorter prompts or fine tuning [..] or [..] caching common answers.}{V167}. 

Another cost driver is constant retraining via fine-tuning of LLMs which becomes even impractical for frequent updates or new datasets. One speaker noted: \xquote{[..] it's not very scalable when there are new documents being added [..]}{V137}.

\subsubsection{\textcolor{lightlimegreen!70!black}{\textbf{Output Verification (11 videos, 6.2\%)}}}\label{para:guardrails}
This theme deals with the verification and validation of LLM outputs, primarily focusing on guardrails and safety mechanisms for safe and reliable interaction with LLMs, such that \xquote{the output [..] is safeguarded against the legal policy role-based and usage-based violations as per organization's policy[..].}{V132}.

Speakers discuss different tools, which work as an intermediary between the user and LLM and take care of the interactions by providing several mechanisms for handling LLM responses, ensuring correctness, safety, and compliance. Besides verifying outputs, these tools may include mechanisms to directly correct LLM output by automatically re-asking the model for correct or compliant output. 

The mentioned benefits of implementing guardrails and safety mechanisms are mostly related to ensure compliance with organizational policies regarding data privacy and protection of intellectual property and is thus often a critical part in the application that allows production deployment.

\subsection{Co-Occurrences of Themes and Topics in Videos}

\begin{figure*}[h]
    \centering
    \includegraphics[width=0.7\linewidth]{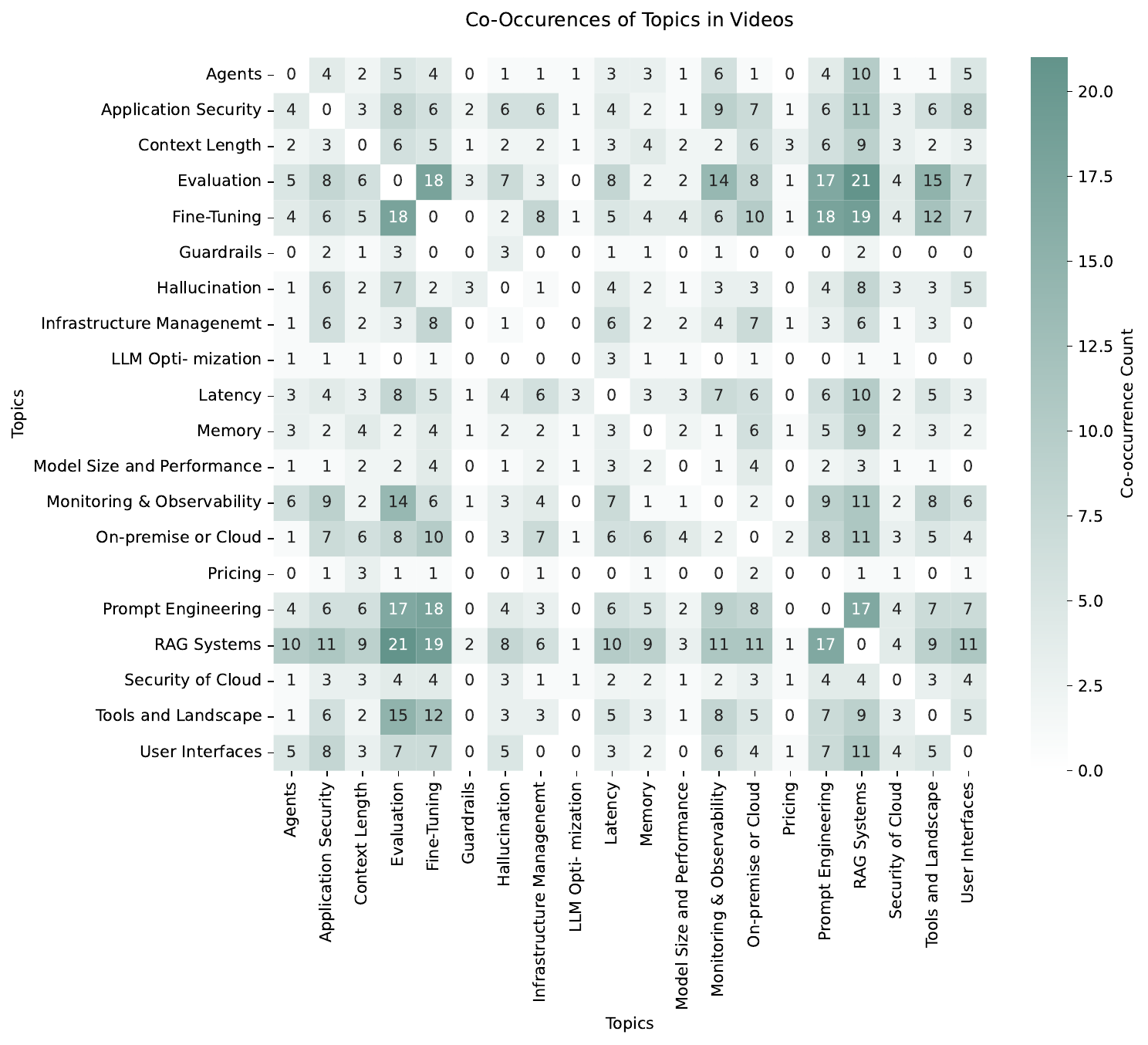}
    \caption{Co-occurrences of topics where darker colors highlight topics occur more often together within individual videos.}
    \label{fig:cooccurences}
\end{figure*}
Our identified topics are not isolated from each other. Clearly, there is an overlap in functionality and themes. Similarly, different topics may represent alternative designs, such that they are mentioned frequently together.
\autoref{fig:cooccurences} shows co-occurrences of topics. The darker a cell, the more often the two corresponding themes occur together. We will now discuss some interesting observations.

\paragraph{Architecture-based Co-Occurrences}
We saw some intersections among topics that fall into the same architecture style. Within a RAG system, prompt engineering is a crucial part (17 joint videos). It specifies how to incorporate the additional context coming from the retrieval part, prepares the LLM for factual answers, and may even rewrite the user prompt to a more suitable phrasing. However, as discussed previously, it is crucial to be able to evaluate the effects of incorporating RAG components. Hence, we observe also that evaluation is often mentioned together with RAG-systems, with in total 21 joint videos---the highest number of co-occurring topics. With similar importance of evaluation for Fine-Tuning (a model; 18) and Prompt Engineering (17).

Some notable further aspects that come with the development of a RAG system are Application Security (11), Context Length (9), Latency (10), Memory (9), Model Size and Performance (11), On-premise or Cloud (11), and User Interfaces (11). So, by just looking at the corresponding row in Figure~\ref{fig:cooccurences}, practitioners can quickly see what further topics may play a role when considering to implement a RAG system.

\paragraph{Alternative Design Choices}
We observe some co-occurrences of topics that may be surprising at first. For example, RAG Systems and Fine-Tuning are not closely related. We may use a fine-tuned model also within a RAG, but the actual reasons why these topics are discussed together are that they are alternative choices for the same problem. Fine-Tuning and RAG systems are two techniques to enable an LLM reason on private, domain-specific, and timely data that was not part of the pre-trained data set. So, the higher number of joint videos (19) indicate such alternative approaches.
Similarly, we see discussions of alternatives between Agents and RAG Systems (10), as well as Prompt Engineering and Fine-Tuning (18).

\paragraph{Tools and Frameworks for Realization}
Finally, joint topics are around available tools and frameworks for specific LLM techniques. That is, following the row Tools and Landscape, we observe joint videos with Evaluation (15; here, tools for obtaining metrics and incorporating evaluation within an LLM inference pipeline are discussed), Fine-Tuning (12; with which tools to fine-tune an LLM), and Monitoring \& Observability (8) and RAG systems (9). So, such an overview can enable a quick delegation to videos talking about concrete tools and frameworks for the corresponding topics. 

From our analysis on co-occurrences of topics, we can infer, at least, three interesting observations: (1) topics to consider when choosing an architecture, (2) alternatives of techniques, and (3) tools for realization. Building upon such a matrix has the potential to become a valuable tool for practitioners in the future for quickly obtaining an overview of the field and find relevant content. Introducing this initial matrix is a notable contribution to the field.

\section{Discussion}
The breadth of the topics is huge. We observe that practitioners exchange architecture styles of LLM-based applications, discuss technical challenges as for fine-tuning and prompt engineering, and highlight very practical aspects, such as risks, ethics, security, and costs. At least for software engineering venues, we do not see such a rich set of topics. Our overview can clearly pinpoint different relevant topics in practice. Notably, the ability to systematically evaluate prompts and RAG systems seem to be one of the biggest concerns, even despite intensive research on that topic~\cite{shankar2024validates, xu2024openp5, simon2024methodology}. It is unclear whether academic approaches are unknown to industry or whether we miss some practical aspects in our research to make an impact. This is an interesting avenue for the future.

There is also limited research on architecture and design of LLM-based systems. While agents have been explored for software engineering tasks ~\cite{suri2023software}, we still miss an overview of architectural decisions and their impacts, specific challenges, and recommendations for practical application. Similarly, while RAG is used for academic approaches, RAG in production settings is largely unexplored. 
Finally, we also observed some similarities and differences between traditional ML and LLMs. For similarities, data collection and cleaning are key for better performance of fine-tuning LLMs; the same is true for training traditional ML models. Furthermore, hyperparameter tuning is often ignored despite its potential~\cite{simon2023exploring}. We find three major differences: (i) stronger dependence of infrastructure during inference, (ii) fuzziness prompt engineering and its sensitivity on results, and (iii) more engineering on model-dependent architecture, such as building RAG systems or multi-agent systems. Based on these observations, it is an important research question to what extent existing knowledge on SE4AI and vice versa can be applied to LLM-based applications.

\subsection{Threats to Validity}

A threat to internal validity may be caused by using BERTopic for the initial automated generation of initial topics. Such an automated process is always a trade-off between scalability and recall, meaning that we can cover substantially more videos with BERTopic, but may have missed some topic. We mitigate the threat of low precision by manually inspecting the paragraphs that have been categorized to a topic. This way, we could also consolidate the topics by removing and merging topics. Still, we might have missed some topics, but by including over 180 videos, we are convinced that the most pressing and frequently appearing topics are obtained.

A threat to external validity or generalization comes from the selection process of the videos to obtain our initial set. Being aware of this possible threat, we applied two search strategies: First, we search for playlists coming from professional communities (e.g., MLOps.community) and industry conference talks about that topic and obtain high quality, industry-relevant material; second, we used keywords to select videos thematically relevant, but without such a community bias. Given the breadth of the topics, we mitigated the threat to external validity.

\section{Conclusion}
In this study, we explored key themes practitioners discuss online regarding the engineering of LLM-based applications. Our semi-automated analysis identified 20 topics, grouped into eight themes, with prominent focus on \emph{Architecture \& Design}, \emph{Model Capabilities \& Techniques}, and \emph{Tools, Infrastructure, \& LLM Providers}. Manual inspection of video transcripts further revealed challenges and recommendations, resulting in a thematic map that offers a valuable reference for practitioners initiating LLM-based projects. By highlighting themes of practical interest, our study points to potential research directions, such as developing specialized hyperparameter tuning mechanisms and examining architectural considerations specific to LLM applications. Overall, this work contributes to bridging industry needs with research opportunities, informing both future practice and academic inquiry in LLM engineering.

\section*{Acknowledgment} The authors acknowledge the financial support by the Federal Ministry of Education and Research of Germany and by Sächsische Staatsministerium für Wissenschaft, Kultur und Tourismus in the programme Center of Excellence for AI-research ``Center for Scalable Data Analytics and Artificial Intelligence Dresden/Leipzig'', project identification number: ScaDS.AI. Siegmund's work has been funded by the German Research Foundation (SI 2171/2-2). This work is supported by the European Social Fund (ESF) and the German state of Saxony under grant number A.100760692 (Project: Teaching-AI).


\bibliographystyle{IEEEtran}
\bibliography{main}
\centering{\subsection*{\scshape{Videos}}}
\begin{footnotesize}
\begin{itemize}
\item[V3] MLOps.community ``Evaluating LLM-based Applications // Josh Tobin // LLMs in Prod Conference Part 2'', \textit{YouTube}, 2023. \url{https://youtube.com/watch?v=r-HUnht-Gns}.
\item[V4] AI Infrastructure Alliance ``Cameron Feenstra – Anzen – Reality Check: Overcoming the Hallucination Hurdle in Generative AI'', \textit{YouTube}, 2023. \url{https://youtube.com/watch?v=CJKth2WROVY}.

\item[V11] MLOps.community ``LLM on K8s // Panel 2 // LLMs in Conference in Production Conference Part 2'', \textit{YouTube}, 2023. \url{https://youtube.com/watch?v=0e5q4zCBtBs}.
\item[V15] Snorkel AI ``How to tune RAG implementations for specialized enterprise tasks'', \textit{YouTube}, 2024. \url{https://youtube.com/watch?v=2vEOfHg0wIk}.
\item[V16] ML Explained - Aggregate Intellect - AI.SCIENCE ``Practical Applications, Impact, and ROI of Generative AI'', \textit{YouTube}, 2023. \url{https://youtube.com/watch?v=68czpakm6dY}.
\item[V17] Amberflo ``Panel Discussion Building and Scaling LLM Applications'', \textit{YouTube}, 2024. \url{https://youtube.com/watch?v=YCRn41SWfPo}.
\item[V19] ML Explained - Aggregate Intellect - AI.SCIENCE ``Running LLMs in Your Environment'', \textit{YouTube}, 2023. \url{https://youtube.com/watch?v=REv-GgieWto}.
\item[V23] Cohere ``Jay Alammar Presents Large Language Models for Real World Applications'', \textit{YouTube}, 2022. \url{https://youtube.com/watch?v=BBQLPu9XqTk}.
\item[V27] MLOps.community ``Guardrails for LLMs: A Practical Approach // Shreya Rajpal // LLMs in Prod Conference Part 2'', \textit{YouTube}, 2023. \url{https://youtube.com/watch?v=e_9o4los7DQ}.
\item[V32] Arize AI ``Practical Tips for Building Production-Grade RAG Applications with LlamaIndex'', \textit{YouTube}, 2023. \url{https://youtube.com/watch?v=TIouyATCHbU}.
\item[V36] Databricks ``Evaluating LLM-based Applications'', \textit{YouTube}, 2023. \url{https://youtube.com/watch?v=2CIIQ5KZWUM}.
\item[V42] LLMOps Space ``Traceability and Observability in Multi-Step LLM Systems  Langfuse  LLMOps'', \textit{YouTube}, 2024. \url{https://youtube.com/watch?v=vRFSVGaNaG4}.
\item[V47] Sam Witteveen ``5 Problems Getting LLM Agents into Production'', \textit{YouTube}, 2024. \url{https://youtube.com/watch?v=06kslWw_QOc}.
\item[V51] AI Infrastructure Alliance ``Hemant Jain – Cohere – Efficiently serving fine-tuned LLMs'', \textit{YouTube}, 2023. \url{https://youtube.com/watch?v=AutBwW8GeDQ}.

\item[V52] Databricks ``Large Language Models in Healthcare: Benchmarks, Applications, and Compliance'', \textit{YouTube}, 2023. \url{https://youtube.com/watch?v=NXN-kMWq6aY}.
\item[V60] AI Infrastructure Alliance ``Amber Roberts–Troubleshooting and Measuring Embedding/Vector Drift for Production Deployments of LMs'', \textit{YouTube}, 2023. \url{https://youtube.com/watch?v=l3GYHKXADW0}.
\item[V63] Snorkel AI ``Q\&A Panel Building GenAI with Your Data'', \textit{YouTube}, 2024. \url{https://youtube.com/watch?v=ncPCk6wF7l0}.
\item[V64] MLOps.community ``LLMs For the Rest of Us // Vikram Sreekanti \& Joseph Gonzalez // LLMs in Prod Conference Part 2'', \textit{YouTube}, 2023. \url{https://youtube.com/watch?v=AaM6nF1a784}.
\item[V66] LlamaIndex ``LlamaIndex Sessions Practical Tips and Tricks for Productionizing RAG (feat. Sisil @ Jasper)'', \textit{YouTube}, 2024. \url{https://youtube.com/watch?v=ZP1F9z-S7T0}.
\item[V68] MLOps.community ``Building Reliable AI Agents // Travis Fischer // LLMs in Production Conference Part 2'', \textit{YouTube}, 2023. \url{https://youtube.com/watch?v=A6wUIOMz7bE}.
\item[V70] FunctionalTV ``LLM Avalanche Panel Build \& Risks'', \textit{YouTube}, 2023. \url{https://youtube.com/watch?v=y61zbIlwnBg}.
\item[V75] MLT Artificial Intelligence ``Large language models in production: opportunities and challenges applying AI to patents (Sam Davis)'', \textit{YouTube}, 2021. \url{https://youtube.com/watch?v=Z1M620vd3u4}.
\item[V96] MLOps.community ``LLM Deployment with NLP Models // Meryem Arik // LLMs in Production Conference Lightning Talk 2'', \textit{YouTube}, 2023. \url{https://youtube.com/watch?v=BN-txmqGxvQ}.
\item[V97] AI Campus Berlin ``Building LLM Applications for Production - AI Campus Berlin'', \textit{YouTube}, 2023. \url{https://youtube.com/watch?v=HyucjMv9_-I}.

\item[V103] MLOps.community ``Challenges and Opportunities in Building Data Science Solutions with LLMs // QuantumBlack Roundtable'', \textit{YouTube}, 2023. \url{https://youtube.com/watch?v=0j0EtPDunyY}.
\item[V105] The TWIML AI Podcast with Sam Charrington ``Building Real-World LLM Products with Fine-Tuning and More with Hamel Husain - 694'', \textit{YouTube}, 2024. \url{https://youtube.com/watch?v=sGqEKzJYrNE}.
\item[V110] FunctionalTV ``LLM Avalanche Panel Enteprise \& Deployment'', \textit{YouTube}, 2023. \url{https://youtube.com/watch?v=K-3i7XeVB9Y}.
\item[V111] MLOps.community ``Ux of a LLM User // Panel 5 // LLMs in Production Conference Part 2'', \textit{YouTube}, 2023. \url{https://youtube.com/watch?v=JISzT9wrsUY}.
\item[V114] ML Explained - Aggregate Intellect - AI.SCIENCE ``Challenges and Solutions for LLMs in Production'', \textit{YouTube}, 2023. \url{https://youtube.com/watch?v=uBxx9VOifCg}.
\item[V126] AI Engineer ``Building Production-Ready RAG Applications Jerry Liu'', \textit{YouTube}, 2023. \url{https://youtube.com/watch?v=TRjq7t2Ms5I}.
\item[V131] MLOps.community ``Vector Databases and Large Language Models // Samuel Partee // LLMs in Production Conference'', \textit{YouTube}, 2023. \url{https://youtube.com/watch?v=GJDN8u3Y-T4}.
\item[V132] MLOps.community ``Transforming AI Safety \& Security // Manojkumar Parmar // LLMs in Production Conference Part 2'', \textit{YouTube}, 2023. \url{https://youtube.com/watch?v=XDvvB-DkmRw}.
\item[V137] Rootcode ``Utilizing Large Language Models in Production - AI Community Meetup - June 2023'', \textit{YouTube}, 2023. \url{https://youtube.com/watch?v=4jFisUYBuGo}.
\item[V141] FunctionalTV ``LLM Avalanche Panel Performance'', \textit{YouTube}, 2023. \url{https://youtube.com/watch?v=qed3heOmGvs}.
\item[V147] AIM Research ``Deploying LLM in production at scale  Anurag Mishra and Puneet Narang  EY Delivery services India'', \textit{YouTube}, 2024. \url{https://youtube.com/watch?v=lBOeELjndJo}.
\item[V156] MLOps.community ``Large Language Models in Production Round-table Conversation'', \textit{YouTube}, 2023. \url{https://youtube.com/watch?v=rpjLTHrl-S4}.
\item[V159] Data Science Dojo ``Emerging Architectures for Large Language Model Applications |  Building a Custom LLM Application'', \textit{YouTube}, 2023. \url{https://youtube.com/watch?v=-eXZhgE1_N4}.

\item[V160] Replit ``Unleashing LLMs in Production: Challenges \& Opportunities. Chip Huyen, Amjad Masad \& Michele Catasta'', \textit{YouTube}, 2023. \url{https://youtube.com/watch?v=ByhMpN2iSbc}.
\item[V161] Anyscale ``Practical Data Considerations for Building Production-Ready LLM Applications'', \textit{YouTube}, 2023. \url{https://youtube.com/watch?v=WLKRAzORtOI}.
\item[V167] DecisionForest ``Biggest Challenge in LLMOps Managing Costs'', \textit{YouTube}, 2023. \url{https://youtube.com/watch?v=R-0XaLsaVNE}.

\item[V174] Weights \& Biases ``Memory in LLM Applications'', \textit{YouTube}, 2023. \url{https://youtube.com/watch?v=3fge-zqZezw}.

\end{itemize}
\end{footnotesize}

\end{document}